\documentclass[twoside,a4paper]{article}
\usepackage{fleqn,epsf,graphicx}
\usepackage{authblk}
\usepackage{verbatim}
\usepackage{graphicx}
\usepackage{geometry}
\usepackage{color}
\usepackage{calc}
\usepackage{epsfig}
\usepackage{amssymb}
\usepackage{amsmath}
\usepackage{cite}
\usepackage{titlesec}

\begin{document}

\title{Luminosity determination for the deutron-deutron reactions using free and quasi-free reactions with WASA-at-COSY detector}

\author{M. Skurzok$^{a}$, P. Moskal$^{a,b}$, W. Krzemie\'n$^{a,c}$\\ 
for the WASA-at-COSY collaboration}

\affil{$^{a}$M. Smoluchowski Institute of Physics, Jagiellonian University, Cracow, Poland}

\affil{$^{b}$Institut f\"ur Kernphysik, Forschungszentrum J\"ulich, Germany}
\affil{$^{c}$National Centre for Nuclear Research, 05-400 Otwock-\'Swierk, Poland}

\maketitle{PACS numbers: 21.85.+d, 21.65.Jk, 25.80.-e, 13.75.-n}


\begin{abstract}
\noindent Two methods of the luminosity determination for the experiment performed by WASA collaboration to search for $^{4}\hspace{-0.03cm}\mbox{He}$-$\eta$ bound state are presented. During the measurement the technique of continous change of the beam momentum in one accelerator cycle (called ramped beam) was applied. This imposes the requirement to determine not only the total integrated luminosity, but also its variation as a function of the beam momentum.
 
\end{abstract}

%
%

\section{Introduction}


\noindent The existence of $\eta$-mesic nuclei in which the $\eta$ meson is bound within a nucleus via the strong
interaction was postulated in 1986 by Haider and Liu \cite{HaiderLiu1}. Since then $\eta$- and $\eta'$-mesic bound states have been searched for in many laboratories~\cite{proc_erice2012,Smyrski,Adlarson_2013,Krzemien_2009, Mersmann, Smyrski_2007, Sokol, Gillitzer, Budzanowski,Nanova2,Tanaka,Afanasiev,Fujioka,Baskov,Krusche_2013, Pheron}.
\noindent Recent theoretical investigations e.g.~\cite{BassTom1,WycechKrzemien,Hirenzaki1,Krusche_Wilkin,Wilkin2014,Nagahiro_2013,Kelkar} support the search for $\eta$ and $\eta'$-mesic bound states, however, so far no firm experimental confirmation of the existence of mesic nuclei has been found.~The discovery of this new kind of an exotic nuclear matter would be very important for better understanding of the $\eta$ and $\eta'$ meson properties and their interaction with nucleons inside nuclear matter~\cite{InoueOset}. Furthermore it would provide information about the $N^{*}$(1535) resonance~\cite{Jido}, as well as about the flavour singlet component of the quark-gluon wave function of the $\eta$ and $\eta'$ mesons~\cite{BassTom}.

\noindent In November 2010 the search for the $^{4}\hspace{-0.03cm}\mbox{He}$-$\eta$ bound state was performed with WASA-at-COSY 
facility~\cite{Adam1} by measuring  
the excitation functions for $dd\rightarrow$ $^{3}\hspace{-0.03cm}\mbox{He} n \pi{}^{0}$ and $dd\rightarrow$ $^{3}\hspace{-0.03cm}\mbox{He} p \pi{}^{-}$ reactions near the $\eta$ production threshold~\cite{mswkpm_epj_2014,wkpmms_acta2014,wkpmms_fbs2014,wkpmjsms_epj2014,mspmwk_actasuppl2013,proc_erice2012}.
The measurement was carried out with a deuteron beam momentum ramping from 2.127 GeV/c to 2.422 GeV/c, corresponding to the range of the excess energy \mbox{Q$\in$(-70,30) MeV}.~During an acceleration process the luminosity could vary due to beam losses caused by the interaction with the target and with the rest gas in the accelerator beam line, as well as due to the changes in the beam-target overlap correlated with momentum variation and adiabatic shrinking of the beamsize. Therefore, it is necessary to determine not only the total integrated luminosity but also its dependence on the excess energy. \\
\noindent The total integrated luminosity is determined based on the $dd\rightarrow$ $^{3}\hspace{-0.03cm}\mbox{He} n$ and quasi free $pp\rightarrow p p$ reactions for which the cross sections were already experimentally established. 
Because of the acceptance variation for the beam momentum range for which $^{3}\hspace{-0.03cm}\mbox{He}$ ions are stopped between two Forward Detector layers, the excess energy dependence of the luminosity is determined based on quasi-free $pp\rightarrow p p$ reaction for which the WASA acceptance is a smooth function of the beam momentum. 

\noindent In this contribution we present the procedure of the calculation of the integrated luminosity and the determination of the luminosity dependence of the excess energy.

\section{Determination of luminosity based on the $\mathbf{dd\rightarrow}\mathbf{^{3}\hspace{-0.03cm}\mbox{He} n}$ \newline reaction}



\noindent The absolute value of the integrated luminosity was determined using the experimental data on the $dd\rightarrow$ $^{3}\hspace{-0.03cm}\mbox{He} n$ cross-sections measured by SATURNE collaboration for four beam momenta in range between 1.65 and 2.49 GeV/c~\cite{Bizard_SATURNE}. The cross section $\sigma_{dd\rightarrow ^{3}\hspace{-0.03cm}{He} n}$ dependence on the transferred momentum squared $t=(\mathbb{P}_{^{3}\hspace{-0.03cm}{He}}-\mathbb{P}_{beam})^{2}$ may be parametrized as follows~\cite{Bizard_SATURNE,Pricking_PhD}:

\begin{equation}
\frac{d\sigma(t-t_{max})}{dt}=\sum_{i=1}^{3} a_{i} e^{b_{i}(t-t_{max})}~\label{eq_1},
\end{equation}


\noindent
where parameters $a_{i}$ and $b_{i}$  are described as a function of the total energy $\sqrt{s_{dd}}$:

\begin{equation}
par_{i}(\sqrt{s_{dd}})=\frac{p_{i}}{\sqrt{s_{dd}}-q_{i}}+r_{i},
\end{equation}

\noindent where the values of $p_{i}$, $q_{i}$ and $r_{i}$ were determined~\cite{Pricking_PhD} by the fit of the above formula to the cross sections measured at SATURNE~\cite{Bizard_SATURNE}.

\noindent Based on the above parametrization we may determine angular dependence of the cross section using a following relation:


\begin{equation}
\frac{d\sigma}{d(cos\theta^{*})}=\frac{d\sigma}{dt} \cdot \frac{dt}{d(cos\theta^{*})}~\label{jeden}
\end{equation}

\noindent
where the Jacobian term $\frac{dt}{d(cos\theta^{*})}=2\cdot |\vec{p}^{\,\,*}_{beam}|\cdot |\vec{p}^{\,\,*}_{^{3}\hspace{-0.05cm}He}|$ is calculated based on the transferred momentum squared in the CM system: 

\begin{equation}
t=(\mathbb{P}_{^{3}\hspace{-0.03cm}{He}}-\mathbb{P}_{beam})^{2}=m_d^2 + m^2_{^{3}\hspace{-0.05cm}He}-2\cdot E^{\,\,*}_{^{3}\hspace{-0.05cm}He}\cdot E^{\,\,*}_{beam}+2\cdot |\vec{p}^{\,\,*}_{beam}|\cdot |\vec{p}^{\,\,*}_{^{3}\hspace{-0.05cm}He}|\cdot cos\theta^{*},\\
\end{equation}

\noindent where $\theta^{*}$ is the $^{3}\hspace{-0.03cm}\mbox{He}$ emission angle in the CM frame.\\

\begin{figure}[h]
\centering
\includegraphics[width=7.0cm,height=5.0cm]{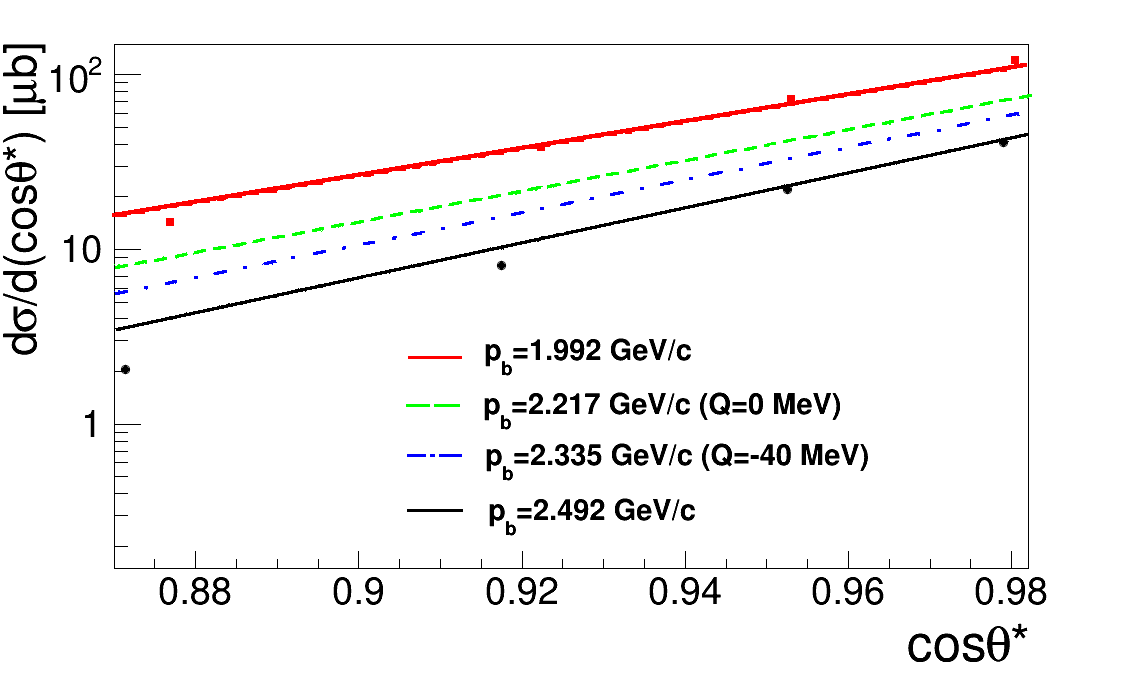} 
\includegraphics[width=7.5cm,height=5.2cm]{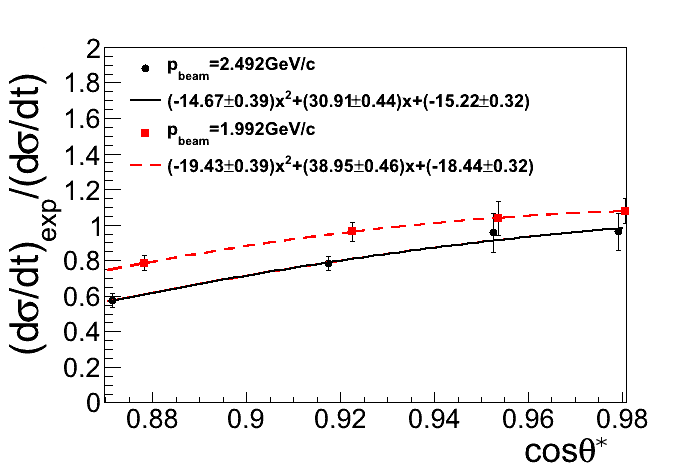} 
\caption{(left) Differential cross section as a function of $cos\theta^{*}$ for SATURNE experimental data (squares/red and dots/black points for fixed beam momentum $p_{beam}$=1.992 GeV/c and $p_{beam}$=2.492 GeV/c, respectively) and obtained from parametrization (top solid/red, dashed/green, dash-dotted/blue, and bottom
solid/black lines for $p_{beam}$ equal to 1.992 GeV/c, 2.217 GeV/c, 2.335 GeV/c and 2.492 GeV/c, respectively). (right) The ratio $\frac{dt}{d(cos\theta^{*})_{exp}}/\frac{dt}{d(cos\theta^{*})}$ for $p_{beam}$=1.992 GeV/c (squares/red) and $p_{beam}$=2.492 GeV/c (dots/black) fitted with second degree polynomial functions (dashed/red and solid/black lines, respectively). The marked errors result from the statistical experimental uncertainties.~\label{fig_Bizzard}}
\end{figure}

\noindent The available experimental data closest to the range of beam momentum used in the WASA-at-COSY experiment for the angular range relevant for our analysis are shown in Fig.~\ref{fig_Bizzard}. Superimposed lines present results of the above described parametrisations for beam momenta the same as experimental points (red and black) and for two exemplary momenta corresponding to Q=0 and Q=-40 MeV.   



\noindent In the angular region of interest the experimental points lie below the curves.
~Therefore, the correction was applied for the $^{3}\hspace{-0.03cm}\mbox{He}$ angular range from about 4$^{\circ}$ to 10$^{\circ}$ which corresponds to the $cos\theta^{*}\in$(0.88,0.98) for the considered reaction. The ratio between experimental \mbox{(SATURNE)} and parametrized cross section $\frac{dt}{d(cos\theta^{*})_{exp}}/\frac{dt}{d(cos\theta^{*})}$ was fitted with second degree polynomial function for both experimental beam momentum values: 1.992~GeV/c and 2.492 GeV/c.~Obtained result is presented in the right panel of Fig.~\ref{fig_Bizzard}. The cross section correction $A$ is calculated for fixed $cos\theta^{*}$ using the fitted functions and interpolated for the proper beam momentum value from range \mbox{$p_{beam}\in(2.127,2.422)$GeV/c}. 
 
\begin{figure}[h!]
\centering
\includegraphics[width=6.0cm,height=4.5cm]{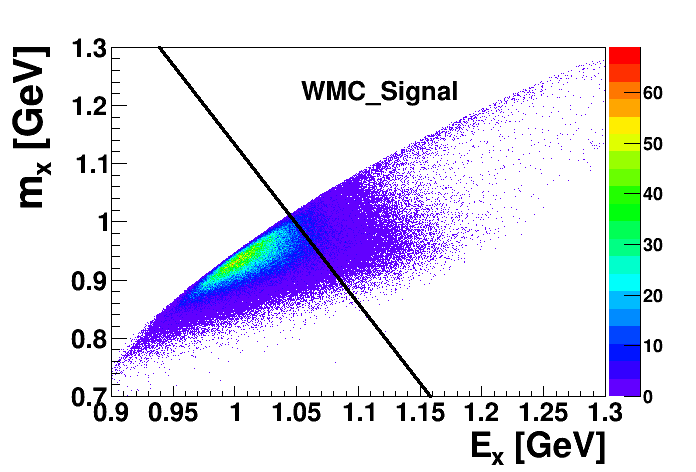}
\includegraphics[width=6.0cm,height=4.5cm]{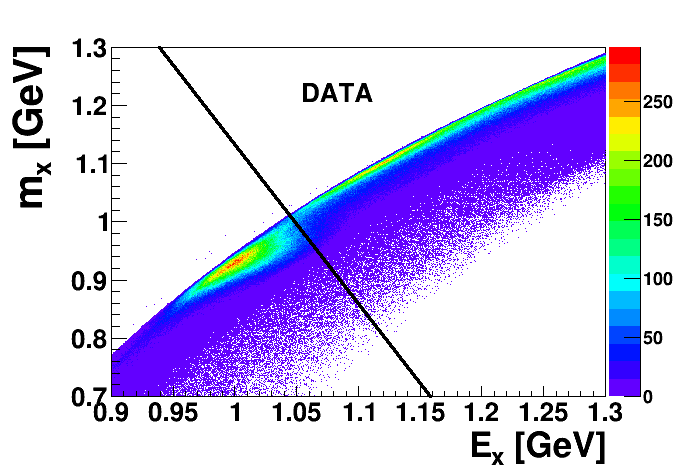} \\
\includegraphics[width=6.0cm,height=4.5cm]{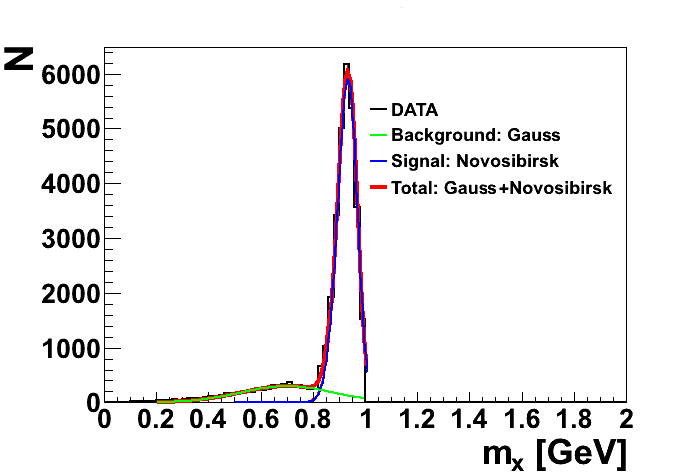}
\caption{(upper panel) The missing mass $m_{x}$ vs. missing energy $E_{x}$ spectrum for simulations (left)
and DATA (right). Applied cut is marked with red line. (lower panel) The missing mass $m_{x}$ spectrum for i. e. $cos\theta^{*}\in$(0.96,0.98) and Q$\in$(0,5) MeV. The red line shows fit to the signal and background while green line shows fit of the Gauss function to the background. Signal peak is marked as a blue line. The main background on the right side of the black line corresponds to the quasi-free \mbox{$dp(n)\rightarrow ^{3}\hspace{-0.03cm}\mbox{He} n \pi^{0}$} reaction.~\label{fig_33}}
\end{figure}

\noindent The measurement of the \mbox{$dd\rightarrow ^{3}\hspace{-0.03cm}\mbox{He} n$} reaction was based on the registration of the outgoing helium in the Forward Detector. Low-energetic $^{3}\hspace{-0.03cm}\mbox{He}$ ions were stopped in the 3rd layer of the Forward Range Hodoscope, while high-energetic ions were stopped in the 4th layer. The helium identification was based on the $\Delta$E-$\Delta$E  
method. The outgoing neutrons were identified using the missing mass technique. In order to reduce background originating from quasi-free \mbox{$dp(n)\rightarrow ^{3}\hspace{-0.03cm}\mbox{He} n \pi^{0}$}, the cut in missing mass $m_{x}$ vs. missing energy $E_{x}$ spectrum was applied as it is presented in upper panel of Fig.~\ref{fig_33}. Additionally, for high beam momentum region background was subtracted via fitting the signal and background function to the missing mass spectrum for different intervals of $cos\theta^{*}$ and beam momentum, what is presented in the lower panel of Fig.~\ref{fig_33}. 

\noindent In order to calculate the total integrated luminosity, the number of events, efficiency, as well as cross section was determined for 5 intervals of $cos\theta^{*}$ in the range from 0.88 to 0.98 and 5 intervals of excess energy Q in the range from -70 MeV to 30 MeV corresponding to the angular range of the reaction and the beam momentum ramping, respectively. The integrated luminosity was then calculated for each $(i,j)$-th interval in following way:\\

\begin{equation}
L^{int}_{i,j}=\frac{N_{i,j}}{\epsilon_{i,j}\cdot\frac{d\sigma_{i,j}}{d(cos\theta^{*})}\cdot \Delta(cos\theta^{*})},
\end{equation} \\

\noindent where $\Delta(cos\theta^{*})$ is the width of the $cos\theta^{*}$ interval. The overall efficiency including reconstruction efficiency and geometrical acceptance of the detector was determined based on the Monte Carlo simulations and is varying between 50\% and 70\%. 

\begin{figure}[h]
\centering
\includegraphics[width=9.0cm,height=6.0cm]
{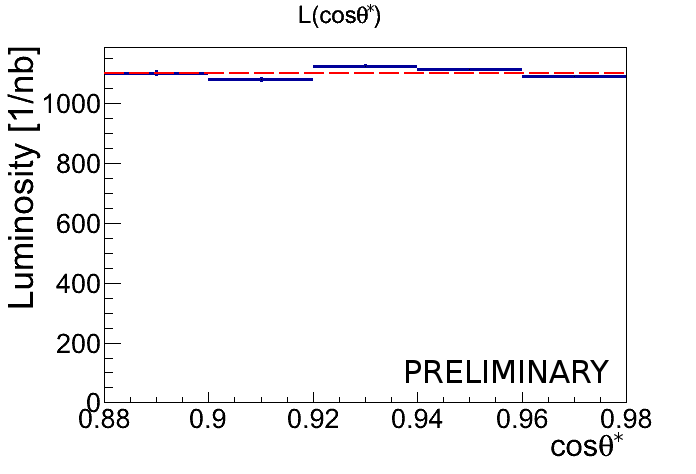}
\vspace{-0.4cm}
\caption{Integrated luminosity as a function of $cos\theta^{*}$.~The statistical uncertainties are marked as a vertical bars. The preliminary established weighted average of integrated luminosity is marked as a dashed red line and is equal to 1102$\pm$2$nb^{-1}$ where only a statistical error is given. The analysis was carried out with condition that the number of "neutral clusters" in  reconstructed in the Central Detector is less than 2.~\label{fig_18}}
\end{figure}

\noindent The preliminary luminosity dependence of $cos\theta^{*}$ for whole excess energy range is presented in Fig.~\ref{fig_18}. The total integrated luminosity was calculated as a weighted average of the luminosities determined for individual $cos\theta^{*}$ intervals:

\begin{equation}
L^{tot}_{dd\rightarrow ^{3}\hspace{-0.03cm}He n}=\frac{\sum_{i=1}^{5} L_{i} \frac{1}{(\Delta L_{i})^2}}{\sum_{i=1}^{5} \frac{1}{(\Delta L_{i})^2}}, \hspace{0.5cm} \Delta L^{tot}_{dd\rightarrow ^{3}\hspace{-0.03cm}He n}=\left(\sum_{i=1}^{5} \frac{1}{(\Delta L_{i})^2}\right)^{-1/2}.\label{Lum_av}\\
\end{equation}\\

\noindent The average integrated luminosity with its statistical uncertainty equals $L^{tot}_{dd\rightarrow ^{3}\hspace{-0.03cm}He n}$=(1102$\pm$2)$nb^{-1}$. It is marked in Fig.~\ref{fig_18} with dashed red line.

\newpage
\section{Luminosity dependence on the excess energy}\label{L_Q}

In order to determine the luminosity dependence on the beam momentum we used the quasi-elastic proton-proton scattering in the deuteron-deuteron collisions: $dd\rightarrow p p n_{sp} n_{sp}$.~In this reaction protons from the deuteron beam are scattered on the protons in the deuteron target. 
We assume that the neutrons are acting only as spectators which means that they do not take part in reactions but move with the Fermi momentum of their parent deuterons. 

\noindent In the case of quasi-free proton-proton scattering the formula for the calculation of the integrated
luminosity can be written in the following form~\cite{Mos_Czyz}:

\begin{equation}
L_{}=\frac{N_{0} N_{exp}}{2\pi \int_{\Delta\Omega(\theta_{lab},\phi_{lab})}\frac{d\sigma}{d\Omega}(\theta^{*},\phi^{*},{p}_{F_{1,2}}, \theta_{F_{1,2}}, \phi_{F_{1,2}})f({p}_{F_{1,2}}, \theta_{F_{1,2}}, \phi_{F_{1,2}})d{p}_{F_{1,2}}dcos\theta_{F_{1,2}}d\phi_{F_{1,2}},d\phi^{*}dcos\theta^{*}}.\label{eq_wzor}\\ 
\end{equation}\\

\noindent The formula is determined based on the fact, that the number of quasi-free scattered protons into the solid angle $\Delta\Omega(\theta_{lab},\phi_{lab})$ is proportional to the integrated luminosity L, as well as the inner product of the differential cross section for scattering into the solid angle around $\theta^{*}$ and $\phi^{*}$ angles expressed in proton-proton CM system: $\frac{d\sigma}{d\Omega}(\theta^{*},\phi^{*},{p}_{F_{1,2}}, \theta_{F_{1,2}}, \phi_{F_{1,2}})$ and the probability density of the Fermi momentum distributions: $f({p}_{F_{1,2}}, \theta_{F_{1,2}}, \phi_{F_{1,2}})$ inside the deuteron beam and deuteron target, respectively.
The detailed description of the luminosity calculation for quasi-free reaction one can find in Ref.~\cite{Mos_Czyz}.

\noindent Due to the complex detection geometry a solid angle corresponding to particular part of the detector cannot be in general expressed in a closed analytical form.~Therefore, the integral in above equation was computed with the Monte-Carlo simulation programme, containing the geometry of WASA detection system and taking into account detection and reconstruction efficiencies.~The Monte-Carlo simulations were carried out for the deuteron beam momentum range $p_{beam}\in$(2.127,2.422)GeV/c corresponding to the experimental ramping.~The program first choose randomly the momentum of the nucleon inside the deuteron beam and deuteron target, respectively, according to the Fermi distribution~\cite{Lacombe}. Then, the total proton-proton invariant mass $\sqrt{s_{pp}}$ and the vector of the center-of-mass velocity are determined. Next, the effective proton beam momentum $p^{prot}_{{beam}}$ was calculated in the frame where one of the proton is at rest and momentum of protons is generated isotropically in the proton-proton center-of-mass frame. Further on, the momenta of outgoing particles are transformed to the laboratory frame and are used as an input in the simulation of the detection system response with the GEANT computing package. For each of $N_{0}$ simulated event we assign a weight corresponding to the differential cross section, which is uniquely determined by the scattering angle and the total proton-proton collision \mbox{energy $\sqrt{s_{pp}}$}.\\
The factor $N_{0}/2\pi$ in Eq.~\ref{eq_wzor} is a normalization constant. It results from the fact that the integral is not dimensionless and its units correspond to the units of the cross sections used for the calculations. Therefore, it must be normalized in such a way that the integral over the full solid angle equals to the total cross section for the elastic scattering averaged over the distribution of the total proton-proton invariant mass $\sqrt{s_{pp}}$ resulting from the Fermi distribution of the target and beam nucleons. In the absence of the Fermi motion it should be simply equal to a total elastic cross section for a given proton beam momentum. A factor $2\pi$ comes from the fact that protons taking part in the scattering are indistinguishable.\\

\begin{figure}[h]
\centering
\includegraphics[width=6.5cm,height=5.0cm]{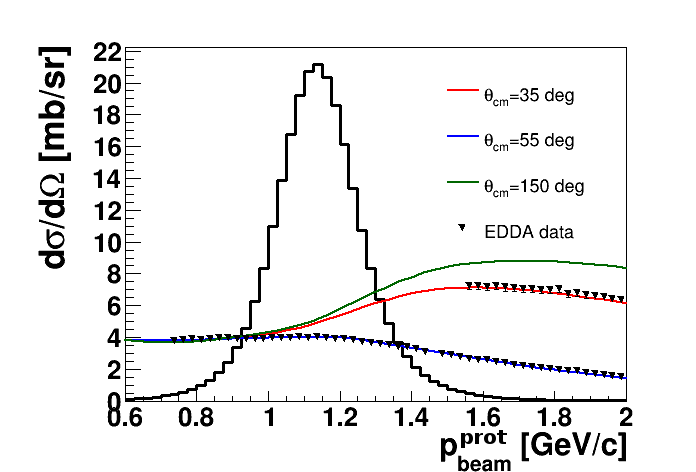}
\includegraphics[width=6.0cm,height=5.0cm]{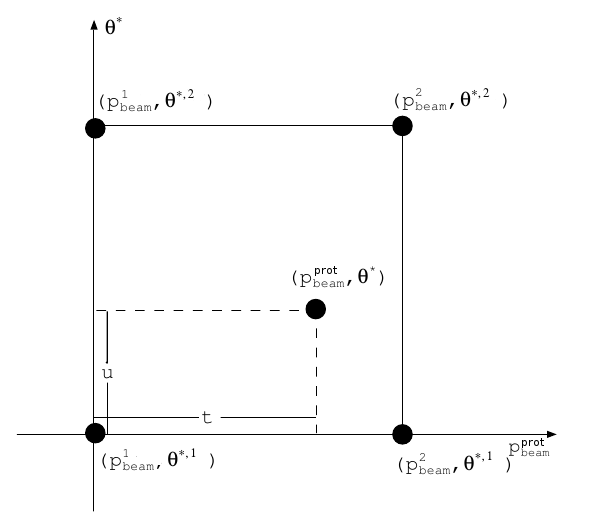} 
\caption{(left) Differential cross sections for proton-proton elastic scattering as a function of the beam momentum for a three values of the scattering angle $\theta^{*}$ in the CM frame. Black points show EDDA collaboration data~\cite{Albers_EDDA}, while lines denote SAID calculations~\cite{SAID_data}. Distribution of the effective beam momentum for quasi-free $pp\rightarrow p p$ reaction calculated for the deuteron beam momentum range $p_{beam}\in$(2.127,2.422)GeV/c is also presented in the figure. (right) Bilinear interpolation of the differential cross section $\frac{d\sigma}{d\Omega}(p^{prot}_{beam},\theta^{*})$. The figure is adapted from~\cite{Mos_Czyz}.~\label{fig_qf_XS}}
\end{figure}

\noindent The differential cross section for quasi free $dd\rightarrow p p n_{sp} n_{sp}$ reaction is a function of the scattering angle $\theta^{*}$ and the total energy in the proton-proton centre-of-mass system $\sqrt{s_{pp}}$ which is dependent on effective proton beam momentum $p^{prot}_{beam}$ seen from the proton in the proton-proton system. In order to calculate it, we have used the cross section values for proton-proton elastic scattering $pp\rightarrow pp$ computed based on the SAID program~\cite{SAID_data} because the EDDA collaboration data base~\cite{Albers_EDDA} is insufficient. The distribution of the effective beam momentum as well as a comparison of the SAID calculations and the existing differential cross section from the EDDA measurements are shown in Fig.~\ref{fig_qf_XS}. As we can see, the differential cross sections calculated using the SAID programme are in agreement with distributions measured by the EDDA collaboration.

\noindent The differential cross section for appropriate $p^{prot}_{beam}$ and $\theta^{*}$ was calculated using bilinear interpolation in the momentum-scattering angle plane according to the formula:

\begin{equation}
\begin{split}
\frac{d\sigma}{d\Omega}(p^{prot}_{beam},\theta^{*})=(1-t)(1-u)\frac{d\sigma}{d\Omega}(p^{1}_{beam},\theta^{*,1})+
t(1-u)\frac{d\sigma}{d\Omega}(p^{2}_{beam},\theta^{*,1})+\\
tu\frac{d\sigma}{d\Omega}(p^{2}_{beam},\theta^{*,2})+(1-t)u\frac{d\sigma}{d\Omega}(p^{1}_{beam},\theta^{*,2})
\end{split}
\end{equation}

\noindent where $t$ and $u$ variables are defined in right panel of Fig.~\ref{fig_qf_XS}.

\noindent The number of experimental events $N_{exp}$ was determined based on conditions and cuts described in details in reference~\cite{Krzemien_PhD}.~In the analysis, at the beginning, we carried out primary events selection applying condition of exactly one charged particle in the Forward Detector (FD) and one particle in the Central Detector (CD). 

\begin{figure}[h]
\centering
\includegraphics[width=8.0cm,height=5.5cm]{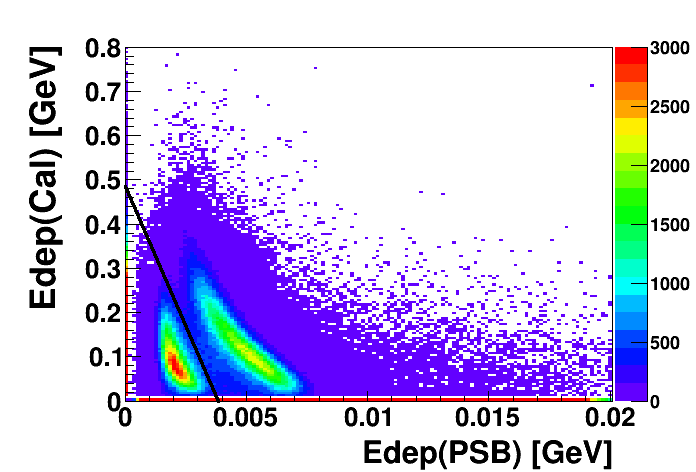} 
\vspace{-0.1cm}
\caption{Experimental spectrum of the energy loss in the Plastic Scintillator Barrel shown as a function of the energy deposited in the Electromagnetic Calorimeter. The applied cut is shown as a black line. Pions in data spectrum are concentrated for Edep(PSB) around 0.003GeV.~\label{fig_psb}}
\end{figure}

\noindent In Ref.~\cite{Krzemien_PhD} we can find detailed studies of the possible background reaction contributions. In case of this analysis the dominating background processes are $dd\rightarrow p p n_{sp} n_{sp}\rightarrow d \pi^{+} n_{sp} n_{sp}$, $dd\rightarrow d_{b} p_{t} n_{sp}$ and $dd\rightarrow p p_{sp} n n_{sp}$, where the subscripts $sp$, $b$ and $t$ denote the spectators, particles from the beam and from the target, respectively. In order to reject the events corresponding to the charged pions registered in the Central Detector, the cut on the energy deposited in the Electromagnetic Calorimeter (Cal) vs. energy deposited in Plastic Scintillator Barrel (PSB) spectrum was applied and is presented in the Fig.~\ref{fig_psb}. 

 
\begin{figure}[h]
\centering
\includegraphics[width=6.0cm,height=4.5cm]{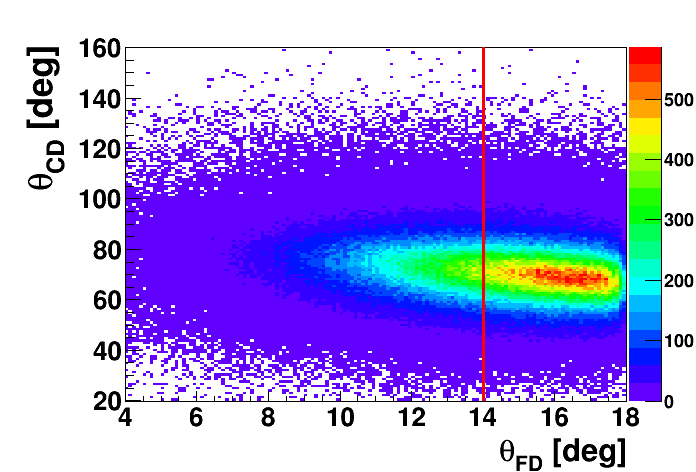}
\includegraphics[width=6.0cm,height=4.5cm]{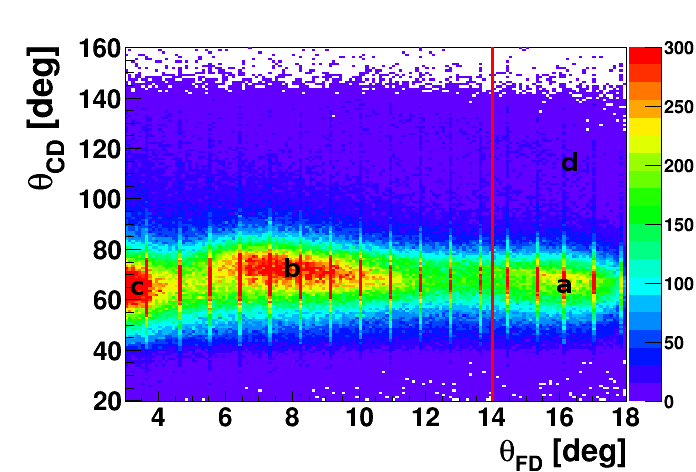} 
\vspace{-0.1cm}
\caption{Correlations between the polar angles $\theta_{FD}$ and $\theta_{CD}$ for the WMC Simulations of $dd\rightarrow p p n_{sp} n_{sp}$ reaction (left panel) and obtained in experiment (right panel).~Applied cut is marked with red line. The indicated area correspond to the: a) $dd\rightarrow p p n_{sp} n_{sp}$, b) $dd\rightarrow d_{b} p_{t} n_{sp}$ and $dd\rightarrow p p n_{sp} n_{sp}$, c) $dd\rightarrow p p_{sp} n n_{sp}$, d) $dd\rightarrow p_{b} d_{t} n_{sp}$.~\label{fig_th_copl}}
\end{figure}

\noindent It is not possible to separate quasi-elastic $p$-$p$ scattering from the quasi-elastic $d$-$p$ scattering, however it was investigated that for the forward scattering angles of about $\theta_{FD}$=17$^{\circ}$, the $d$-$p$ cross sections are about 20 times smaller than $p$-$p$ cross sections and we take this uncertainty of abut 5\% as a systematic error to the final result. The applied cut in polar angle $\theta_{FD}$ is shown in Fig.~\ref{fig_th_copl}. In order to subtract the background coming from $dd\rightarrow p_{b} d_{t} n_{sp}$ reaction, the range $\theta_{CD}\in$(40,100)$^{\circ}$ was taken into account in further analysis.    
  
  
  
\begin{figure}[h!]
\centering
\includegraphics[width=8.0cm,height=5.5cm]{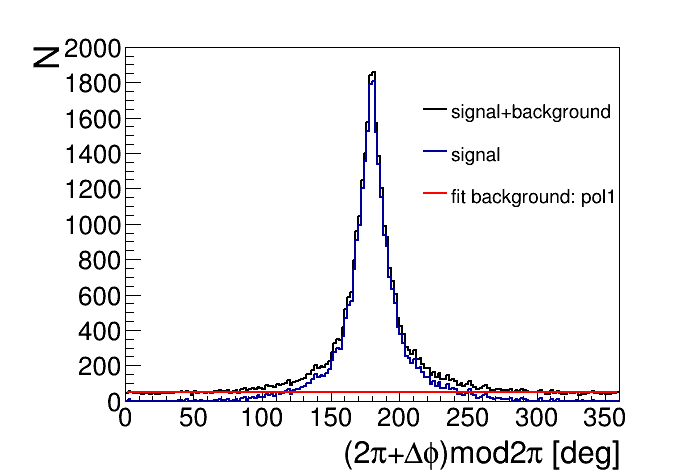}\\
\caption{Distributions of (2$\pi+\Delta\phi$)mod2$\pi$, where $\Delta\phi=\phi_{CD}-\phi_{FD}$ is the difference of azimuthal angles in Central Detector and Forward Detector.~The example spectrum for one of the Q intervals (black line) with marked fit function (red line) and signal peak after background subtraction (blue line) is presented.~\label{fig_copl}}
\end{figure}

\noindent Additionally, the background was subtracted in $\Delta\phi=\phi_{FD}-\phi_{CD}$ spectrum. In order to symetrize the background instead of $|\Delta\phi|$ we define (2$\pi+\Delta\phi$)mod2$\pi$. Afterwards, the background was fitted with 1st order polynomial for each of excess energy Q intervals.~The exemplary \mbox{(2$\pi+\Delta\phi$)mod2$\pi$} spectrum is presented in Fig.~\ref{fig_copl}.

\vspace{-0.4cm}

\noindent After all cuts and conditions described above, the number of experimental data was determined and the luminosity was calculated according to formula (\ref{eq_wzor}) for each excess energy interval taking into account also prescaling factor of the applied experimental trigger equal to 4000 as well as shadowing effect equal to 9\%.~The latter results from the fact that proton is shadowed by the neutron inside the deuteron which reduces the probability of the quasi-elastic scattering. Unfortunately, there are no experimental results about the shadowing in $dd\rightarrow p p n_{sp} n_{sp}$ collisions. However, we can try to estimate it based on the probability that a neutron shadows the proton in one deuteron which equals 0.045~\cite{Chiavasa} and assume that shadowing appears independently in deuteron beam and deuteron target. The rough estimation of the probability that the shadowing will not take place in $dd$ reaction (1 - 0.045)$^{2}$ gives about 0.91.

\begin{figure}[h]
\centering
\includegraphics[width=11.0cm,height=7.0cm]{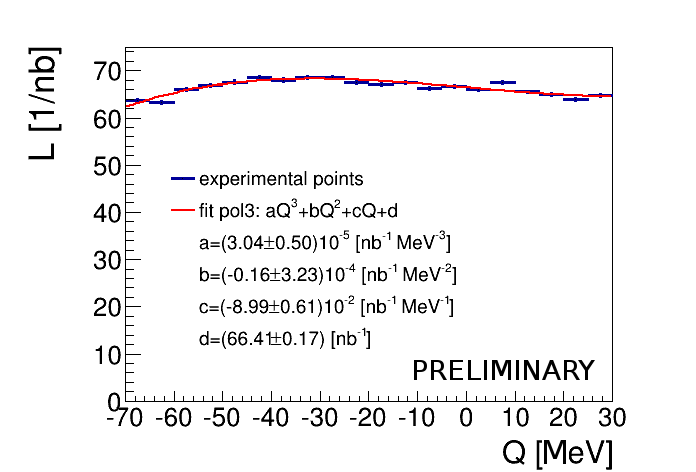}
\caption{Integrated luminosity calculated for experimental data for quasi-free $dd\rightarrow p p n_{sp} n_{sp}$ reaction (blue points) with fitted 3 degree polynomial function (red line).~\label{lum_norm_fit}}
\end{figure}

\noindent The preliminary result is presented in Fig.~\ref{lum_norm_fit}. The statistical uncertainty of each point is about 1\%. The luminosity variation (increase in the excess energy range from about -70MeV to -40MeV, and then decrease) is caused by the change of the beam-target overlapping during the acceleration cycle and also by adiabatic beam size shrinking~\cite{Lorentz}. The obtained total integrated luminosity within its statistical uncertainty is equal to $L^{tot}_{dd\rightarrow p p n_{sp} n_{sp}}$=(1329$\pm$2)$nb^{-1}$. For further analysis the luminosity was fitted by 3 degree polynomial $aQ^{3}+bQ^{2}+cQ+d$. The fitted function is marked with the red line in Fig.~\ref{lum_norm_fit}.

\section{Systematics}

\noindent In case of the $dd\rightarrow$ $^{3}\hspace{-0.03cm}\mbox{He} n$ reaction one source of the systematic error originates from the variation of the cuts used for separation of high-energetic helium in Forward Detector and is equal to about 2\%. Additionally we have also taken into account an uncertainty due to the method used for the background subtraction amounting to 1.6\%. Another source of the luminosity calculation error is connected to normalization to SATURNE experiment and originates from three independent sources: i) statistical error of the SATURNE data (6.5\%), ii) normalization uncertainty of the SATURNE data for the $dd\rightarrow$ $^{3}\hspace{-0.03cm}\mbox{He} n$ cross sections (7\%) and iii) assumption of linear interpolation between SATURNE points used for the estimation of the correction for the parametrized cross section presented in Fig.~\ref{fig_Bizzard} ($<$1.8\%). 

\noindent The systematical errors for $dd\rightarrow p p n_{sp} n_{sp}$ analysis resulting from the change of the cuts used for the separation of the quasi-free $pp$ scattering from the background (Fig. \ref{fig_psb} and Fig. \ref{fig_th_copl}) is equal to to about 4.1\%. The another contribution to the systematical error comes from the assumption of the potential model of the nucleon bound inside the deuteron and is equal to about 0.8\%. This uncertainty was established as the difference between results determined using the Paris~\cite{Lacombe} and the CDBonn~\cite{CDBonn} potentials. The next source of the systematic error may be attached to the assumption of the bilinear approximation of the cross section shown in Fig.~\ref{fig_qf_XS} (right).~This systematical uncertainty was estimated using assumption in which instead of the interpolation we took the cross section value from the closest data point in the effective proton beam momentum - scattering angle plane. The performed calculations give the difference of about 1.8\%. Additionally we have also taken into account an uncertainty related to the background subtraction in (2$\pi+\Delta\phi$)mod2$\pi$ spectra which does not exceed 0.6\%. The systematical uncertainty includes also contribution connected to the shadowing effect. Untill now, we have no theoretical estimation of the possible error of this effect calculation, therefore conservatively we take as an systematic uncetainty half of this effect: 4.5\%. In the systematical error calculation we also take into account the uncertainty 5\% resulting from the background of the quasi-elastic $d$-$p$ scattering (Sec.~\ref{L_Q}). The normalization error includes two contributions: normalization uncertainty of the EDDA data (4\%) and the systematical errors for $pp$ elastic scattering cross-sections (2.7\%)~\cite{Albers_EDDA}. The cross section was approximated by the calculation using the SAID procedure. Because, the SAID cross section very well describes EDDA data, we assume the systematical errors of the differential cross section based on EDDA calculations.\\ 

\noindent The total integrated luminosity calculated based on $dd\rightarrow$ $ ^{3}\hspace{-0.03cm}\mbox{He} n$ and the quasi-free $dd\rightarrow p p n_{sp} n_{sp}$ reactions with statistical, systematical and normalization error are equal to $L^{tot}_{dd\rightarrow ^{3}\hspace{-0.03cm}He n}=(1102\pm2_{stat}\pm28_{syst}\pm107_{norm})nb^{-1}$ and $L^{tot}_{dd\rightarrow p p n_{sp} n_{sp}}=(1329\pm2_{stat}\pm108_{syst}\pm64_{norm})nb^{-1}$, respectively. The systematical and normalization errors were calculated by adding in quadrature the appropriate contributions described above.

\section{Summary}

\noindent We carried out the luminosity determination for the experiment performed with WASA-at-COSY to search for the $^{4}\hspace{-0.03cm}\mbox{He}$-$\eta$ bound states in deuteron-deuteron fusion. The luminosity was calculated based on two reactions: $dd\rightarrow$ $ ^{3}\hspace{-0.03cm}\mbox{He} n$ and the quasi-free $dd\rightarrow p p n_{sp} n_{sp}$.~We calculated the total average integrated luminosity and compared it for both channels. The obtained results are consistent, however within large normalization errors. 

\section{Acknowledgements}

\noindent We acknowledge support by the Foundation for Polish Science - MPD program, co-financed by the European
Union within the European Regional Development Fund, by the Polish National Science Center through grant No. 2011/01/B/ST2/00431, Grant PRELUDIUM No 2013/11/N/ST2/04152 and by the FFE grants of the Research Center J\"ulich.


\end{document}